\documentclass{PoS}

\title{Charmonium, $D_s$ and $D_s^*$ from overlap fermion on domain wall fermion configurations
\thanks{This work is supported in part by the National Science Foundation of China (NSFC) under
Grants No.10835002, No.11075167, No.11105153 and also by the U.S. DOE Grants No. DE-FG05-84ER40154.
Y. C. and Z. L. also acknowledge the support of NSFC and DFG (CRC110).}}

\ShortTitle{Charmonium, $D_s$ and $D_s^*$ from overlap fermion on DWF configurations}

\author{\speaker{Y.B. Yang},$^{1,2}$ Y. Chen,$^1$
A. Alexandru,$^3$ S.J. Dong,$^2$
   T. Draper,$^2$ M. Gong,$^{1,2}$ F.X. Lee,$^3$ A. Li,$^{4}$ K.F. Liu,$^2$ Z. Liu, $^1$ M. Lujan, $^3$ and N. Mathur $^5$ 

\begin{center}   
   ($\chi$QCD  Collaboration)
\end{center}
   \\
        $^1$ Institute of High Energy Physics and Theoretical Physics Center for Science Facilities, \\
        Chinese Academy of Sciences, Beijing 100049, China\\
        E-mail: \email{ybyang@ihep.ac.cn}, \email{cheny@ihep.ac.cn},
        \email{liuzf@ihep.ac.cn},\\
        $^2$ Department of Physics and Astronomy, University of Kentucky, Lexington, KY 40506, USA\\
        E-mail: \email{liu@pa.uky.edu}\\
        $^3$ Department of Physics, George Washington University, Washington, DC 20052, USA\\
        $^4$ Institute for Nuclear Theory, University of Washington, Seattle, WA 98195, USA \\
        $^5$ Department of Theoretical Physics, Tata Institute of Fundamental Research, Mumbai 40005, India}

\abstract{We take a new approach to determine the scale parameter $r_0$, the physical masses of
strange and charm quarks through a global fit which incorporates continuum extrapolation, 
chiral extrapolation and quark mass interpolation to the lattice data. The charmonium and charm-strange
meson spectrum are calculated with overlap valence quarks on $2+1$-flavor domain-wall fermion
gauge configurations generated by the RBC and UKQCD Collaboration. We use the masses of $D_s$, $D_s^*$
and $J/\psi$ as inputs and obtain $m_c^{\overline{\rm MS}}(2\,{\rm GeV})=1.110(24)\,{\rm GeV}$,
$m_s^{\overline{\rm MS}}(2\,{\rm GeV})=0.104(9)\,{\rm GeV}$ and $r_0=0.458(11)\,{\rm fm}$.
Subsequently, the hyperfine-splitting of charmonium and $f_{D_s}$ are predicted to be
$112(5)\,{\rm MeV}$ and $254(5)\,{\rm MeV}$, respectively.}

\FullConference{31st International Symposium on Lattice Field Theory - LATTICE 2013\\
        July 29 - August 3, 2013\\
        Mainz, Germany}

\begin{document}
\section{Introduction}

In lattice QCD simulations, quark masses and the strong coupling constant are bare parameters
in the QCD action. The coupling constant is related closely to the dimensional quantity on
the lattice, the lattice spacing $a$, which should be determined first through a proper
scheme to set the scale. After that, the physical point is approached through extrapolation or interpolation of the
bare quark masses by requiring that the physical values of hadron masses and other physical
quantities are reproduced. In this proceeding, we implement a global fit scheme combining the
continuum limit extrapolation, the chiral extrapolation and the quark mass interpolation to set the
scale parameter $r_0$ and determine quark masses (at a specific energy scale $\mu$) simultaneously.
We adopt the overlap fermion action for valence quarks and carry out the practical calculation on
the $2+1$-flavor domain wall fermion gauge configurations generated by the RBC/UKQCD
Collaboration~\cite{Allton:2008dy,Aoki:2010dy}. It has been verified that the charm quark region
can be reached by the overlap fermions on the six ensembles of the RBC/UKQCD configurations at two
lattice spacings $a\sim 1.7$ GeV and $a\sim 2.3$ GeV~\cite{Li:2010pw}, such that by applying the
multi-mass algorithm in the calculation of overlap fermion propagators, we can calculate the hadron
spectrum and other physical quantities at quite a lot of quark masses ranging from the chiral
region ($u,d$ quark mass region) to the charm quark region. This permits us to perform a very
precise interpolation to the physical strange and charm quark masses. In
order to compare with experimental values, the lattice values of dimensionful physical
quantities should be converted to the values in physical units through a scale parameter, for which
we choose the Sommer's parameter $r_0$~\cite{Sommer:1994r0} and express it in the units of lattice 
spacing at each
ensemble through the calculation of static potential. As such, all the dimensionful quantities
calculated on the lattice (and their experimental values) can be expressed in unit of $r_0$. In
order for the physical quantities calculated at different bare quark masses and different lattice
spacings to be fitted together, we calculate the quark mass renormalization constant to convert the
bare quark masses to the renormalized quark masses at a fixed energy scale in the $\overline{\rm MS}$
scheme . With these prescriptions, the  physical quark masses and the $r_0$ can be determined by
some physical inputs, through which we can predict other quantities at the physical point in the
continuum limit.

\section{Quark mass renormalization and the scale setting }
\begin{figure}[Htbp]
\begin{center}
  \includegraphics[scale=0.45]{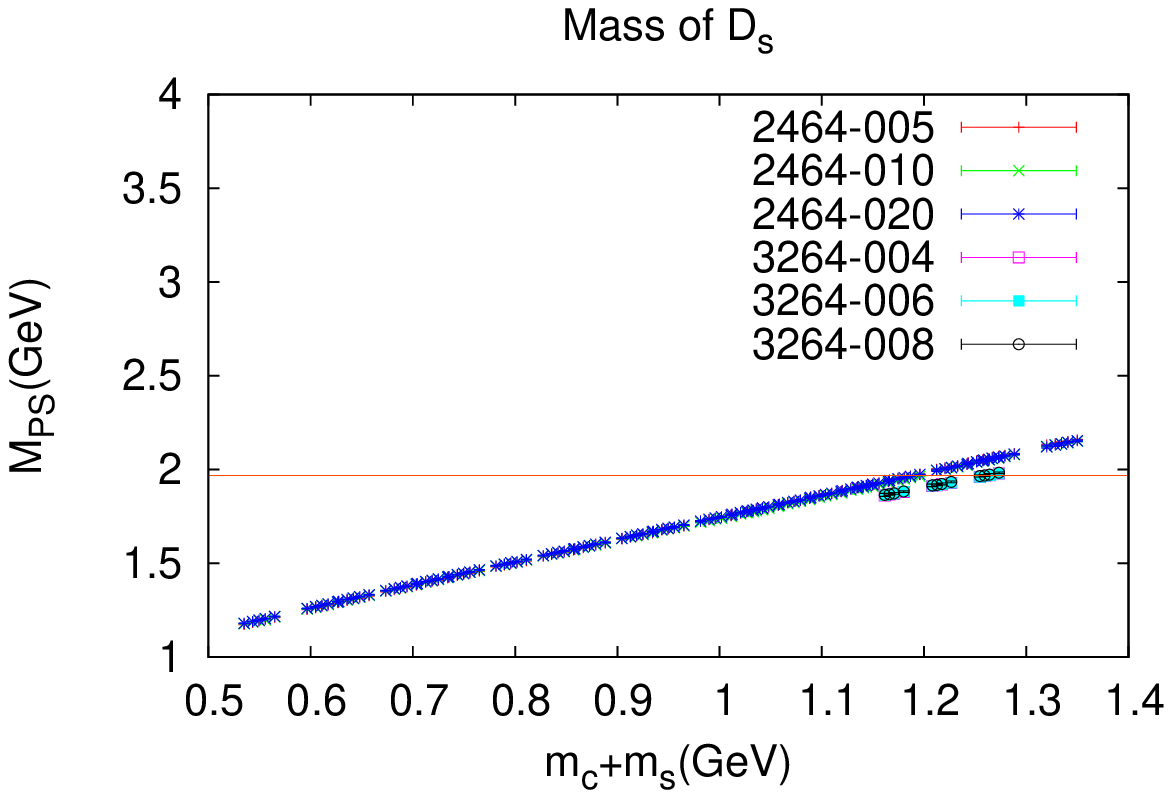}
  \includegraphics[scale=0.45]{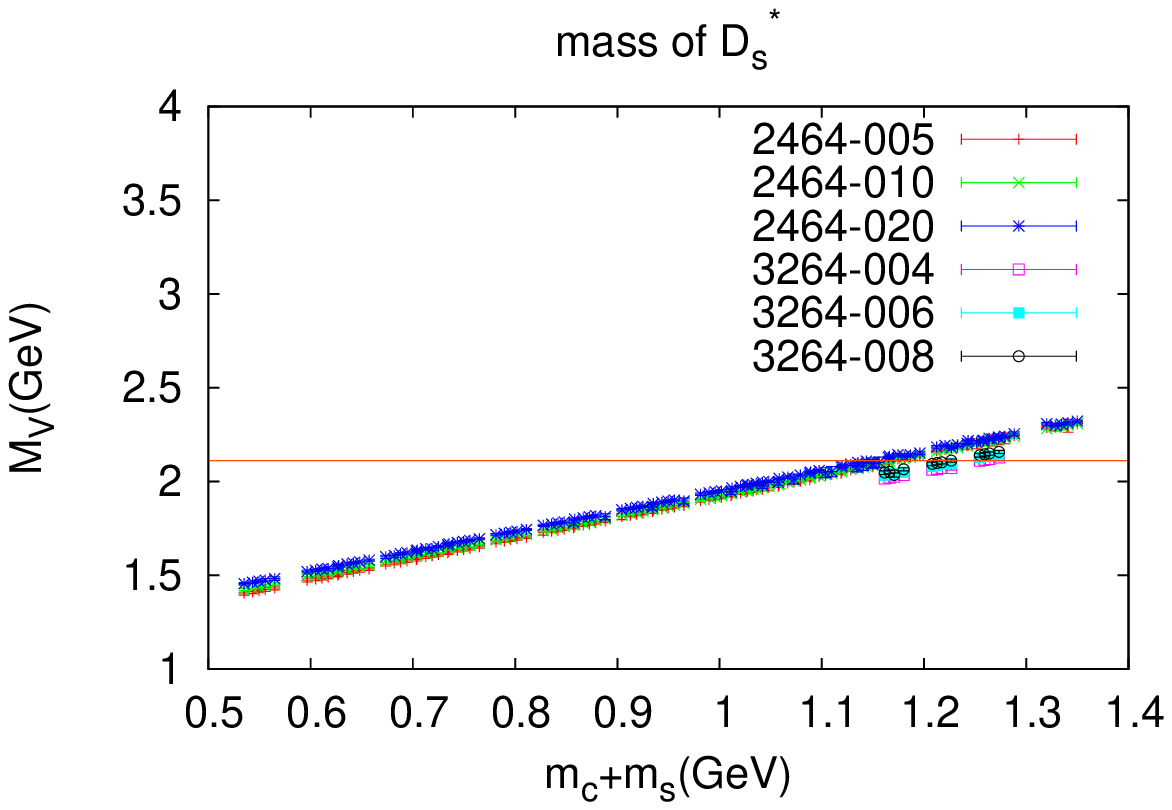}
  \includegraphics[scale=0.45]{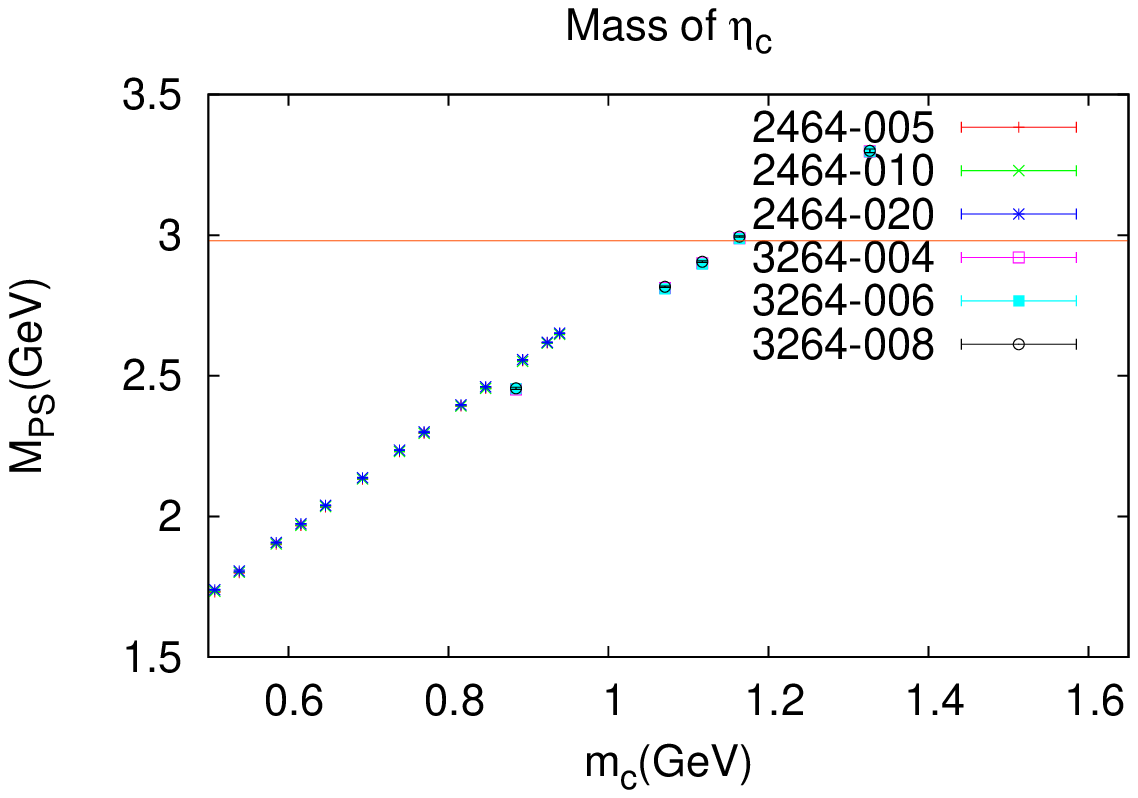}
  \includegraphics[scale=0.45]{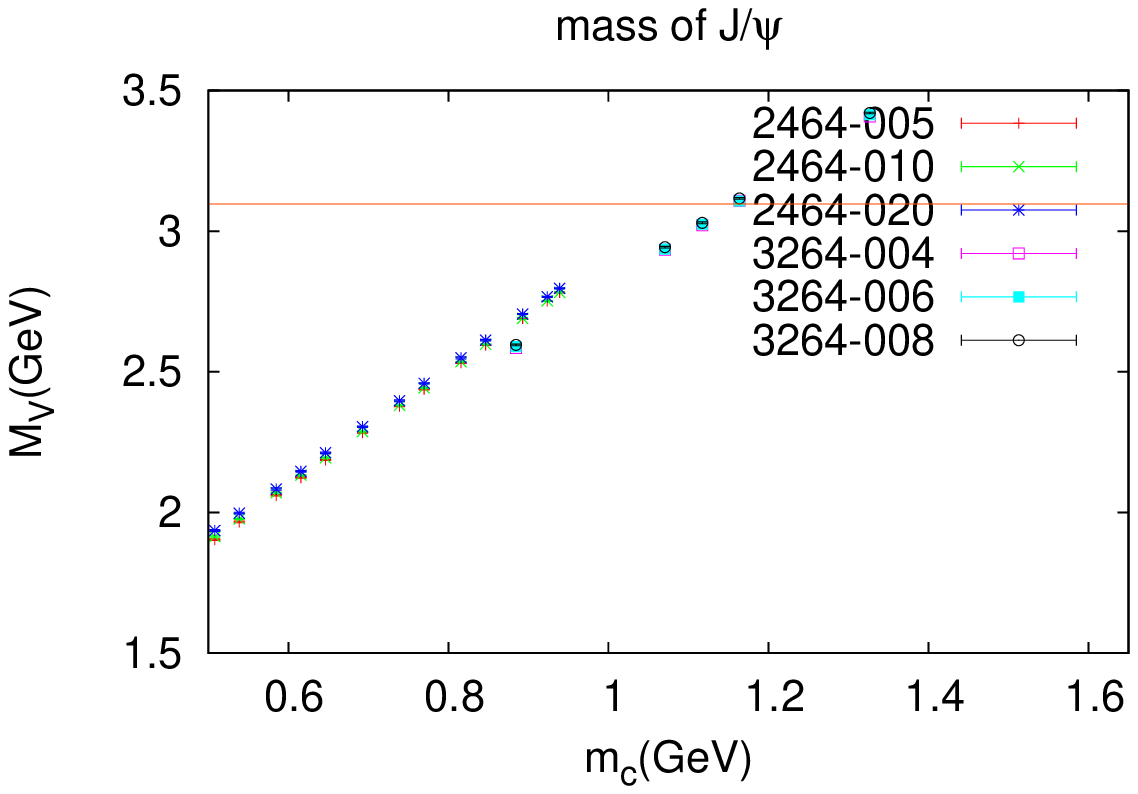}
\caption{\label{fig:swaves}The quark mass dependence of the masses of $D_s$, $D_s^*$, $\eta_c$, and
$J/\psi$ are illustrated in the plots for the six RBC/UKQCD configuration ensembles, where the
linear behaviors in $m_c a$ or $m_ca+m_sa$ are clearly seen.}
\end{center}
\end{figure}

\begin{figure}[Htbp]
\begin{center}
\includegraphics[scale=0.5]{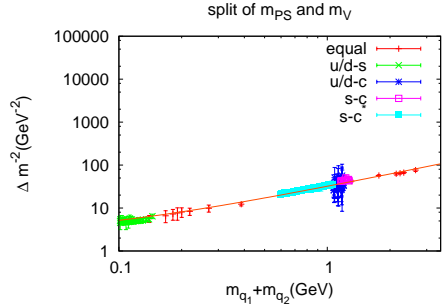}
\caption{The quark mass dependence of the hyperfine splittings $\Delta m=m_V-m_{\rm {PS}}$ for
$\bar{c}s$ and $\bar{c}{c}$ systems. For clarity, the $(\Delta m)^{-2}$ versus $m_{q_1}+m_{q_2}$
are plotted in the figure, where one can see an almost linear behavior throughout the range
$m_{q_1}+m_{q_2}\in [0.1,3]\,{\rm GeV}$. This behavior suggests the relation $\Delta
m\sim\frac{1}{\sqrt{m_{q_1}+m_{q_2}+\delta }}$.}\label{fig:splitting}
\end{center}
\end{figure}

Our calculation are carried out on the $2+1$ flavor domain wall fermion configurations generated by
the RBC/UKQCD collaboration with the parameters listed in Tab.~\ref{table:para}. 
\begin{table}[bh]
\begin{center}
\begin{tabular}{ccccccc}\hline\hline
$\beta$ & $L^3\times T$  & $m_s^{s}a$ &   &  $m_l^{s}a$  &    &  $m_{\rm res}a $ \\
\hline
2.13    &$24^3\times 64$ & 0.04   &0.005&0.01    &0.02  &  0.00315(4)      \\
2.25    &$32^3\times 64$ & 0.03   &0.004&0.006   &0.008 & 0.00067(1) \\
\hline\hline
\end{tabular}
\caption{\label{table:para}The parameters for the RBC/UKQCD configurations~\cite{Aoki:2010dy}.
$m_s^{s}a$ and $m_l^{s}a$ are the mass parameters of the strange sea quark and the light sea quark,
respectively. $m_{\rm res}a$ is the residual mass of the domain wall sea quarks.}
\end{center}
\end{table}
For valence quarks we use the overlap fermion operator $D_{\rm ov}=1+\gamma_5\epsilon(H_W(\rho))$ to define
the effective massive fermion operator
\begin{equation}
D_c(ma)\equiv \frac{\rho D_{\rm ov}}{1-D_{\rm ov}/2}+ma,
\end{equation}
where $H_W(\rho)=\gamma_5 D_W(\rho)$ with $D_W(\rho)$ the Wilson-Dirac operator with a negative
mass parameter $-\rho$, and the parameter $ma$ is purely the bare current quark mass (in
lattice units) and free of additive renormalization owing to the good chiral property $\{\gamma_5,
D_c(0)\}=0$. Through the multi-mass algorithm, quark propagators $S_F(ma)=D_c^{-1}(ma)$ for dozens of different valence quark masses $ma$ can be calculated simultaneously, such that we can
calculate multiple physical quantities at each valence quark mass and obtain clear
observation of the quark mass dependence of these quantities. We estimate the physical strange
quark mass to be around $m_s a = 0.056$ for the $24^3\times 64$ lattice and $m_s a = 0.039$ for the
$32^3\times 64$ lattice, therefore we choose the $m_sa$ to vary in the range $m_s a \in
[0.0576,0.077]$ and $m_s a \in [0.039, 0.047]$ for the two lattices, respectively. 
Figure~\ref{fig:swaves} shows an almost linear bare quark mass dependence of the masses of $J/\psi$,
$\eta_c$, $D_s$, and $D_s^*$ calculated from the six gauge ensembles. In addition to the above
observation, we also find that the hyperfine splitting of vector and pseudoscalar mesons,
$\Delta=M_V-M_{PS}$, varies with quark masses like $\propto 1/\sqrt{m}$~\cite{Li:2010pw}, as shown in
Fig.~\ref{fig:splitting}. Taking into account the sea quark mass dependence, $\Delta$ can be well
described by the formula
\begin{equation}
M_{V}-M_{PS}=\frac{C+C_1 m_l}{\sqrt{m_{q_1}+m_{q_2}+\delta m}}
\end{equation}
to a high precision. The detailed discussion of this dependence will be presented in Ref.~\cite{yang:2013sl}. 
Finally, the global fit formula for the meson system is
\begin{eqnarray}\label{eq:para-cs-ori}
M_{meson}&=& (A_0 +A_1 m_c +A_2 m_s +A_3 m_l
         + (A_4+A_5 m_l)\frac{1}{\sqrt{m_c+m_s+\delta}})\nonumber\\
   &\times& \big(1+ B_0 a^2 +B_1 m_c^2 a^2 + B_2 m_c^4a^4\big)
   + A_4 a^2
\end{eqnarray}
with $\delta$ a constant parameter. Note that $A_{2}$ is set to zero for the charm quark-antiquark
system, and $A_1$ is expected to be close to 1(or 2) for the meson masses of $\bar{c}s$($\bar{c}c$)
system. We keep the $m_ca$ correction to the forth order and add explicit $O(a^2)$ correction term
to account the artifacts due to the gauge action and other possible artifacts. There should be
$m_sa$ corrections, but they are very small and are neglected.


In lattice QCD, the bare quark masses are input parameters in lattice units,
say, $m_q a$. However in the global fit including the continuum extrapolation using
Eq.~(\ref{eq:para-cs-ori}), one has to convert the $m_qa$ to the renormalized current quark mass
$m_q^{R}(\mu)$ at a fixed scale $\mu$ which appears uniformly in Eq.~(\ref{eq:para-cs-ori}) for
different lattice spacings. This requires two issues to be settled beforehand.
The first is the renormalization constant $Z_m$ of the quark mass for a fixed lattice spacing $a$.
Since we use the overlap fermion operator $D_{\rm ov}$, if the quark field $\psi$ is replaced by
the chirally regulated field $\hat{\psi}=(1-\frac{1}{2}D_{\rm ov})\psi$ in the definition of the
interpolation fields and the currents, it is expected that there are relations $Z_S=Z_P$ and
$Z_V=Z_A$, where $Z_S$, $Z_P$, $Z_V$, and $Z_A$ are the renormalization constants of scalar,
pseudoscalar, vector, and axial vector currents, respectively. In addition,
$Z_m$ can be derived from $Z_S$ by the relation $Z_m=Z_S^{-1}$. In the
calculation of $Z_S$ and other $Z$'s of the quark bilinear currents, we adopt the RI-MOM scheme to
do the non-perturbative renormalization on the lattice first, then convert them to the
$\overline{\rm MS}$ scheme using ratios from continuum perturbation theory (The numerical details
can be found in Ref.~\cite{Liu:2013ucx}). The relations between $Z$'s
mentioned above are verified, and $Z_S$ at the scale $\mu=2$GeV in the
$\overline{\rm MS}$ is determined to be $Z_S^{\overline{\rm MS}}(2\,{\rm GeV})=1.127(9)$ for the
$24^3\times 64$ lattice and $ 1.056(6)$ for the $32^3\times 64$ lattice. Besides the statistical
error, systematic errors from the scheme matching and the running of quark masses in the $\overline{\rm MS}$
scheme are also considered in Ref.~\cite{Liu:2013ucx}). The systematic error from the running quark mass in the $\overline{\rm MS}$ scheme
is negligible small, while the one from scheme matching is at four loops, and has
a size of about 1.4\%. 

The second is the precise determination of the lattice spacing $a$. This is a subtle question
because $a$ has direct relationship with the coupling constant $\beta$ and the sea quark mass. On
the other hand, the continuum limits of physical quantities are independent of $a$. So we need
a dimensionful physical quantity as a scale parameter, which is independent of quark masses, such that
dimensional quantities such as hadron masses, lattice spacing $a$, and quark masses can be
expressed in units of this parameter. A proper choice for this purpose is Sommer's scale
parameter $r_0$. In each gauge ensemble, the ratio $r_0/a=\sqrt{(1.65-e_c)/(\sigma a^2)}$ can be
determined very precisely through the derivation of the static potential $V(r/a)$, where $e_c$ and
$\sigma a^2$ are the parameters in $V(r/a)$, say, $aV(r/a)= aV_0 -e_c /(r/a) + \sigma a^2 (r/a)$.
Table~\ref{table:r0} lists the $r_0/a$'s for the six ensembles we are using. $r_0/a$ has obvious
$m_la$ dependence, so we extrapolate it to the chiral limit $\tilde{m}_l=m_la+m_{\rm res}a=0$ by
a linear fit, $r_0(\tilde{m}_la,a)/a=r_0(0,a)/a+ c \tilde{m}_la $ ,for the ensembles with a same
$\beta$. The extrapolated values $C(a)\equiv r_0(\tilde{m}_la,a)/a$ are also listed in Tab.~\ref{table:r0}
\begin{table}[Htbp]
\begin{center}
\begin{tabular}{c|cccc|ccccc}\hline\hline
       &           &$\beta=2.13$ &           &          &        & $\beta=2.25$   &       &
       \\\hline
$m_la$ & 0.02      &  0.01       & 0.005     &   0      & 0.008  & 0.006          & 0.004 & 0 \\
$r_0/a$& 3.906(3)  & 3.994(3)    & 4.052(3)  & 4.126(11)& 5.421(5)& 5.438(6) & 5.459(6)   &
5.504(4) \\
\hline\hline
\end{tabular}
\caption{\label{table:r0}
$r_0/a$'s for the six ensembles we are using.
}
\end{center}
\end{table}

With the above prescriptions, we can replace the renormalized quark masses and $a$ by the bare
quark mass parameters $m_q a$, $C(a)$, $r_0$, and $Z_m(2\,{\rm GeV},a)$ as
\begin{equation}
m_q^R(2\,{\rm GeV}) = Z_m(2\,{\rm GeV},a) (m_qa) \frac{C(a)}{r_0}
\end{equation}
In this way $r_0$ enters into Eq.~(\ref{eq:para-cs-ori}) as a new parameter and can be fitted
simultaneously with the parameters $A_i$ and $B_i$ in Eq. (2.3).

In addition to the $1S$ and $1P$ charmonium masses and $D_s/D_s^*$ masses, we also obtain
predictions of the decay constant of $D_s$, namely, $f_{D_s}$. $f_{PS}$ is defined as
\begin{equation}
Z_A \langle 0|\bar{\psi}_a \gamma_4\gamma_5\psi_b|PS\rangle = E_{PS}f_{PS},
\end{equation}
where $E_{PS}$ is the energy of the pseudoscalar and $Z_A$ is the renormalization constant of the
axial vector current, or alternatively through the PCAC relation
\begin{equation}
\langle 0|\bar{\psi}_a\gamma_5\psi_b|PS\rangle = \frac{E_{PS}^2}{m_{q_a}+m_{q_b}}f_{PS}.
\end{equation}
The later is obviously renormalization independent since $Z_m Z_P=1$ for the overlap fermion. With
the $Z_A$ calculated in Ref.~\cite{Liu:2013ucx}, we find the two definitions are compatible with each
other within errors.

In practice, we carry out a correlated global fit to the following quantities in the six ensembles -- the masses of $J/\psi$,
$\eta_c$, $\chi_{c0}$, $\chi_{c1}$, $h_c$, $D_s$ and  $D_s^*$ mesons and $f_{D_s}$ with jackknife
covariance matrices. As a first step, we obtain the parameters $A_i$'s and $B_i$'s for each
quantity. Subsequently we use the experimental values $m_{J/\psi}=3.097\,{\rm GeV}$,
$m_{D_s}=1.968\,{\rm GeV}$, and $m_{D_s^*}=2.112\,{\rm GeV}$ to determine $m_c^{\overline{\rm
MS}}(2{\rm GeV})$, $m_s^{\overline{\rm MS}}(2{\rm GeV})$, and $r_0$. Finally, we use these physical
parameters to predict the masses of $\eta_c$, $\chi_{c0}$, $\chi_{c1}$, $h_c$, and $f_{D_s}$ at the physical pion point and with $O(a^2)$ corrections. The
results are illustrated in Tab.~\ref{table:results}.
\begin{table}[tbp]
\begin{center}
\begin{tabular}{c|ccc|ccccc}
\hline\hline
&$r_0$ &   $m_s^{\overline{\rm MS}}(2{\rm GeV})$   & $m_c^{\overline{\rm MS}}(2{\rm GeV})$ &$m_{J/\psi}-m_{\eta_c}$  & $m_{\chi_{c0}}$   & $m{\chi_{c1}}$   & $m_{h_{c}}$ & $f_{D_s}$\\
&(fm)  &    (GeV)              & (GeV)             & (GeV)                   & (GeV)  & (GeV) &
(GeV) & (GeV)\\
\hline
 PDG      &   --  & 0.095(5)  & 1.09(3)    & 0.117(1)               & 3.415           & 3.511            & 3.525      & 0.258(6)\\
 this work& 0.458& 0.104     & 1.110       &  0.1119                  & 3.411                 & 3.498    &    3.518  & 0.2542\\
 \hline
 $\sigma$(\rm stat)&0.011&0.006&0.012        &  0.0054                  & 0.035                 & 0.045    &    0.028  & 0.0049\\
 $\sigma$($r_0$/a)&0.002&0.000&0.000           &  0.0000                  & 0.003                 & 0.000    &    0.000  & 0.0000\\
 $\sigma$($\frac{\partial{r_0}}{\partial{a^2}}$)
                &0.003&0.004&0.005   &  0.0005                 & 0.003                 & 0.006    &    0.002  & 0.0003\\
 $\sigma$(MR/stat)   &--&0.003 &0.020        &  0.0000                  & 0.003                 & 0.011    &    0.004  & 0.0003 \\
 $\sigma$(MR/sys)   &--&0.000 &0.003        &  0.0000                  & 0.000                 & 0.000    &    0.000  & 0.0000 \\
$\sigma$(all)   &0.011&0.009&0.024      &  0.0054                  & 0.046                & 0.047    &    0.028   & 0.0049\\
\hline\hline
\end{tabular}
\caption{The final results from the global fit in this work. The physical masses of
$D_s$, $D_s^*$, and $J/\psi$ are used as inputs to determine $r_0$, $m_s^{\overline{\rm
MS}}(2{\rm GeV})$ and $m_c^{\overline{\rm MS}}(2{\rm GeV})$. The physical masses of the $1P$ charmonia can
be reproduced with these values. The hyperfine splitting $m_{J/\psi}-m_{\eta_c}$ and the decay constant
$f_{D_s}$ are also predicted. The error budget is given for each quantity with 
$\sigma({\rm stat})$, the statistical error, and two kinds of systematic errors, 
one from $r_0$ and 
another one from mass renormalization.
For the one from $r_0$, 
it includes
  $\sigma(r_0)$ the uncertainty owing to the statistical error of
$r_0/a$ on each ensemble, and $\sigma(\frac{\partial{r_0}}{\partial(a^2)})$ the systematic uncertainty
from the $a$ dependence of $r_0/a$. For the one from mass renormalization, 
it includes $\sigma(MR/stat)$ from the statistical uncertainties of $Z_m$ in the RI/MOM scheme which are 
not correlated on the two lattices, and $\sigma(MR/sys)$ from the systematic uncertainty in matching from RI/MOM to $\overline{MS}$ which is the same on the two lattices and only contributes to the uncertainty of quark masses. The total error $\sigma(all)$
combines all the uncertainties in quadrature. \label{table:results}}
\end{center}
\end{table}
Here is the description of the error budget: i) The statistical errors are the jackknife errors from
the global fit. ii) The systematic uncertainty due to the linear $a^2$ continuum extrapolation
cannot be controlled at present since we have only two lattice spacings. iii) For the chiral extrapolation
we only use the linear fit in the $u,d$ sea quark mass and have not considered a sophisticated
fit based on chiral perturbation theory, so this systematic uncertainty has not been analyzed. iv)
We consider two possible systematic uncertainties introduced by $r_0$, one of which is from the
statistical error of $C(a)=r_0(a)/a$ (denoted as $\sigma(r_0/a)$), and the other is from non-zero
$a^2$ dependence of $r_0(a)$ (denoted as $\sigma(\frac{\partial{r_0}}{\partial{a^2}})$). {\it The
latter is not so straightforward. In our global fit, we take the $r_0$ at a finite lattice
spacing to be the one in the continuum limit.
To address the issue that our prediction
for $r_0$ (0.458(11) fm) is slightly smaller than the one from RBC/UKQCD (0.48(1) fm), 
we set
$\frac{\partial{r_0}}{\partial{a^2}}=0.2$ (which makes the $r_0(a)$ at the two lattice spacings to be around
0.48 as the RBC-UKQCD collaboration so determines)
, so as to check the changes of the predictions to estimate their
systematic errors. It turns out, the $\chi^2$ of the global fit is insensitive to this
dependence and the results do not change much.}  v) The error $\sigma({\rm MR})$ takes into account
two kinds of uncertainties of $Z_m(\mu)$, one of which is the statistical error of $Z_m(\mu)$ in the RI/MOM scheme, and the
other is due to the systematic error of perturbative matching and the running of the $\overline{\rm
MS}$ masses to the scale of 2 GeV. All of these uncertainties are combined together in quadrature to
give the total uncertainty $\sigma({\rm all})$ of each physical quantity.


\begin{figure}[tbh]
\begin{center}
\includegraphics[scale=0.6]{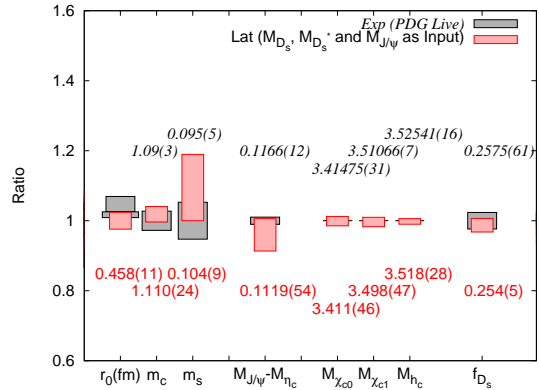}
\caption{The ratios between PDG Live averages and our simulations. Note that the numbers of
PDG Live are in italic type, and all the numbers are in unit of GeV, except $r_0$. For $r_0$,
we list the one from HPQCD (0.4661(38)fm) and RBC-UKQCD (0.48(1)fm) for
reference.}\label{fig:ratio}
\end{center}
\end{figure}
\section{Summary}
With overlap fermions as valence quarks on domain wall fermion configurations generated by the
RBC/UKQCD Collaboration, we have undertaken a global fit scheme combining the chiral extrapolation, the
physical quark mass interpolation, and the continuum extrapolation. We use the physical masses of
$J/\psi$, $D_s$, and $D_s^*$ as the inputs to determine $r_0$, the charm and strange quark masses to
be
\begin{equation}
     r_0=0.458(11)\,{\rm fm},
~~~~~m_{s}^{\overline{\rm MS}}({\rm 2GeV})=0.104(9)\,{\rm GeV}, ~~~~~m_{c}^{\overline{\rm MS}}({\rm
2GeV})=1.110(24)\,{\rm GeV}
\end{equation}
Our $r_0$ is smaller than $0.48(1)$ fm obtained by RBC/UKQCD~\cite{Arthur:2012opa} but
close to the HPQCD result 0.4661(38) fm~\cite{Davies:2009tsa}. With these results, we can reproduce
the physical masses of $\chi_{c0}$, $\chi_{c1}$ and $h_c$, and further predict the hyperfine
splitting $ m_{J/\psi}- m_{\eta_c}$ and the decay constant $f_{D_s}$,
\begin{equation}
m_{J/\psi}- m_{\eta_c}=112(5)\,{\rm MeV},~~~~~f_{D_s}=254(5)\,{\rm MeV}.
\end{equation}
The errors we quote above are quadratic combinations of the statistical and systematic errors.

\end{document}